\documentclass[aps,prc,twocolumn,times,graphicx,tighten,showpacs]{revtex4}
\usepackage{graphicx}
\usepackage[dvips]{color}

\begin{document}

\title{Meson-exchange currents and quasielastic antineutrino cross sections 
in the SuperScaling Approximation}
\author{J.E. Amaro}
\affiliation{Departamento de F\'{\i}sica At\'{o}mica, Molecular y Nuclear
and Instituto Carlos I de F\'{\i}sica Teorica y Computacional, Universidad de Granada,
  18071 Granada, SPAIN}
\author{M.B. Barbaro}
\affiliation{Dipartimento di Fisica Teorica, Universit\`a di Torino and
  INFN, Sezione di Torino, Via P. Giuria 1, 10125 Torino, ITALY}
\author{J.A. Caballero}
\affiliation{Departamento de F\'{\i}sica At\'{o}mica, Molecular y Nuclear,
Universidad de Sevilla,
  41080 Sevilla, SPAIN}
\author{T.W. Donnelly}
\affiliation{Center for Theoretical Physics, Laboratory for Nuclear
  Science and Department of Physics, Massachusetts Institute of Technology,
  Cambridge, MA 02139, USA}

\begin{abstract}
We evaluate quasielastic double-differential antineutrino cross
sections obtained in a phenomenological model based on the
superscaling behavior of electron scattering data and estimate the
contribution of the vector meson-exchange currents in the 2p-2h
sector. We show that the impact of meson-exchange currents for
charge-changing antineutrino reactions is much larger than in the
neutrino case.
\end{abstract}



\maketitle

The recent MiniBooNE data on muon neutrino charged-current quasielastic (CCQE)
scattering~\cite{AguilarArevalo:2010zc} have raised an important
debate on the role played by both nuclear and nucleonic ingredients 
in the description of the reaction.
Unexpectedly, the cross section turns out to be substantially underestimated by
the Relativistic Fermi Gas (RFG) model, unless 
an unusually large {\em ad hoc} value of the axial mass $M_A\simeq 1.35$ GeV/c$^2$ 
(as compared with the standard value $M_A\simeq 1$ GeV/c$^2$) is employed
in the dipole parametrization of the nucleon axial form factor. From 
comparisons with electron scattering data the RFG model is known, however, 
to be too crude to account for the nuclear dynamics --- therefore this
result is perhaps more an indication of the incompleteness of the 
theoretical description of the nuclear many-body problem  
than an unambiguous indication of a larger axial mass. 

At the level of the impulse approximation (IA), a number of more
sophisticated descriptions of the nuclear dynamics other than the RFG
also underpredict the measured CCQE cross section (see, {\it e.g.,}
\cite{Leitner:2006ww,Martini:2009uj,Benhar:2010nx,Amaro:2010sd,Amaro:2011qb,Nieves:2011pp}).
Possible explanations of this puzzle have been proposed, based either
on multinucleon knockout \cite{Martini:2009uj,Nieves:2011yp} or on
particular treatments of final-state interactions through
phenomenological optical potentials~\cite{Meucci:2011vd}, indicating
that contributions beyond the naive starting point play an important
role in QE neutrino reactions.

In \cite{Amaro:2010sd} the predictions of the SuperScaling Approach
(SuSA) model including 2p-2h Meson-Exchange Currents (MEC) model were
presented and compared with the MiniBooNE data on muon neutrino
scattering from $^{12}$C. The SuSA approach has been discussed at
length in
\cite{Donnelly:1998xg,Donnelly:1999sw,Maieron:2001it,Amaro:2004bs}:
comparisons with inclusive electron scattering data have been made
using scaling functions obtained from the {\em longitudinal} inclusive
response which is then employed universally for the QE transverse
scaling function (and hence the transverse inclusive QE response) and
for the inelastic response (production of $\pi$, $\Delta$, DIS, {\it
  etc.}).  This representation of the QE and inelastic parts of the
inclusive cross section is then added to the 2p-2h MEC contribution
obtained in a relativistic model~\cite{DePace:2003xu,Amaro:2010ph} 
to yield the
total inclusive cross section. Good agreement is found for comparison
of this representation with experiment for a wide range of
kinematics. The SuSA approach is then to use exactly the same scaling
functions to obtain the QE and 2p-2h MEC contributions for the
so-called CCQE cross section, actually the sum of what is called QE in
studies of electron scattering plus what is called MEC to distinguish
the latter from the contributions that are predominantly due to
single-nucleon ejection via one-body operators.

The results of the purely SuSA model, {\it i.e.} based on the assumed
universality of the scaling function for electromagnetic and weak
interactions, for the double-differential, single-differential and
total CCQE neutrino cross sections have been shown to fall below the
data for most of the experimental angle and energy bins.  The
inclusion of 2p-2h MEC in the SuSA approach yields larger cross
sections and accordingly better agreement with the data, although
theory still lies below the data at larger angles where the cross
sections are smaller~\cite{Amaro:2010sd}.

\begin{figure}[ht]
\label{fig:nubar}
\includegraphics[scale=0.6, bb=90 150 469 730]{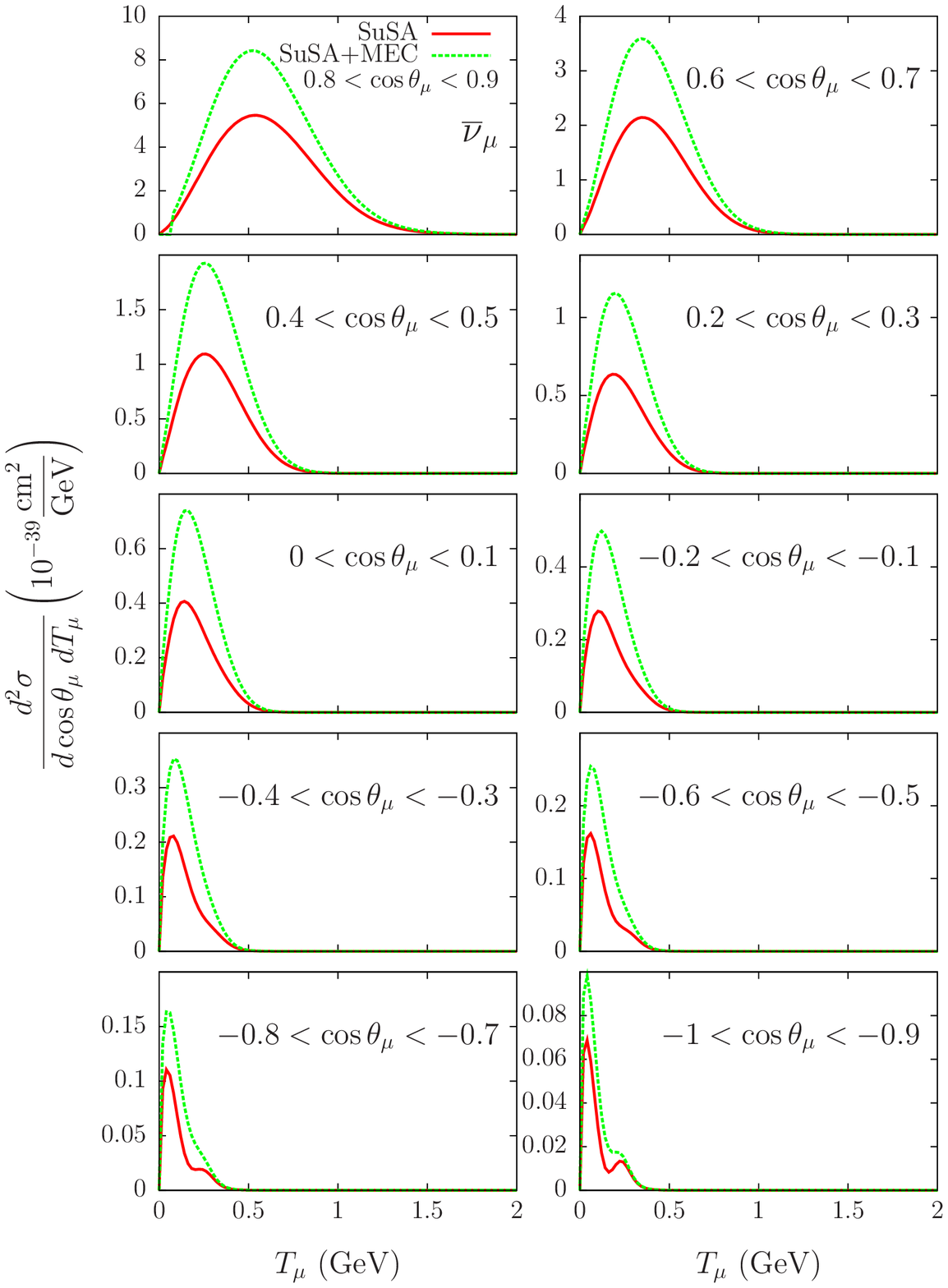}
\caption{ (Color online) Flux-integrated double-differential cross
  section per target nucleon for the $\overline\nu_\mu$ CCQE process
  on $^{12}$C displayed versus the $\mu^+$ kinetic energy $T_\mu$ for
  various bins of $\cos\theta_\mu$. The lower curves in each panel
  (red) show the SuSA QE results, while the upper curves (green) show
  those results plus the contributions from 2p2h MEC.}
\end{figure}
%
\begin{figure}[ht]
\label{fig:nu}
\includegraphics[scale=0.6, bb=90 150 469 730]{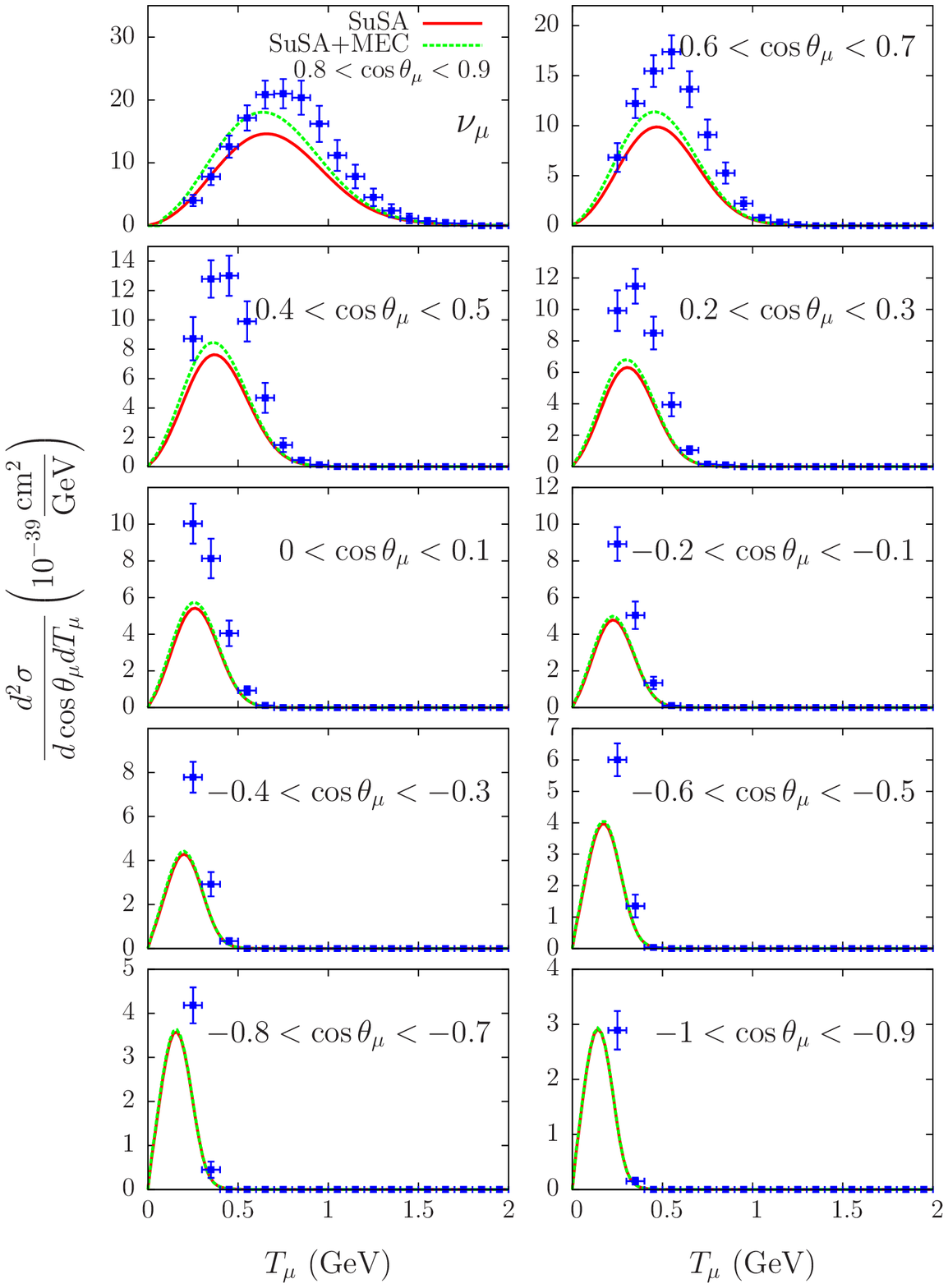}
\caption{ (Color online) As for Fig.1, but for $\nu_\mu$
  scattering versus $\mu^-$ kinetic energy $T_\mu$. The data are from
  Ref.~\cite{AguilarArevalo:2010zc}.}
\end{figure}
In this letter we apply the same model to antineutrino scattering. 
The results are shown in Figs.~1-4.
In particular, in 
Fig.~1
antineutrino CCQE cross sections integrated over the $\overline\nu_\mu$
MiniBooNE flux~\cite{AguilarArevalo:2008yp}
are shown as
functions of $\mu^+$ kinetic energies $T_\mu$ for angular bins $\cos
\theta_\mu$ ranging from forward to backward angles.  For each panel
in the figure two curves are shown, the lower in each case (red) being
the basic SuSA QE result and the upper (green) being this plus the
contributions from 2p-2h MEC in the transverse vector part of the
cross section. For comparison, in Fig.~2
the neutrino cross sections versus $\mu^-$ kinetic
energies are shown for the same kinematical conditions, together with
the data from \cite{AguilarArevalo:2010zc}. The results here have
already been reported in~\cite{Amaro:2010sd,Amaro:2011qb}, although here the full range of kinematics is presented.

The effects from 2p-2h MEC for the antineutrino case are especially
striking: for instance, for the first angular bin shown ($0.8 < \cos
\theta_\mu < 0.9$) the MEC contribute about 38\% to the total cross
section at the peak, while for $0 < \cos \theta_\mu < 0.1$ this rises
to about 44\% at the peak. In contrast (see Fig.~2)
the relative percentage coming from MEC is much
smaller in the neutrino case. The origin of these differences is
clear. For neutrinos one has three basic contributions, namely, from
the SuSA QE contributions to the transverse and longitudinal VV and
transverse AA responses together with small contributions from the
charge-longitudinal AA responses, contributions from 2p2h MEC (which
in leading order only enter in the transverse VV sector) and the VA
interference contribution. For neutrinos the last is constructive and
the MEC effects are to be weighed against relatively large QE
contributions. However, for antineutrinos the VA interference is
destructive; accordingly the total QE contribution is significantly
reduced for antineutrinos and consequently when the MEC are added they
play a much more significant role. To illustrate this point, consider
the angular bin $0.6 < \cos \theta_\mu < 0.7$: while the neutrino and
antineutrino fluxes are different, making the comparisons not quite
trivial, if one evaluates the three contributions at the peaks of the
cross sections one is led to conclude that the QE contribution without
the VA interference is about 6.0, the VA interference is about 3.9 and
the MEC about 1.5 in the units used for the figures. So for neutrinos
the total QE result is about 9.9, while the total QE result for
antineutrinos is about 2.1. Clearly the MEC contribution of 1.5 in
these units has a much greater impact for antineutrinos than for
neutrinos. Indeed, measurements of both reactions would likely provide
a stringent test of the modeling of the 2p-2h MEC.  Moreover, given
measurements using different nuclear targets, where one expects the
relative contributions from QE scattering and MEC to differ in going
from nucleus to nucleus, one would have an additional way to probe the
two types of response. To make these observations more quantitative,
in Fig.~3 the ratios with and without MEC are shown for the 10 panels
used in Figs.~1 and 2, illustrating the much larger role of MEC in the
antineutrino case. Note that the small second peak that shows up at
larger scattering angles (most apparent in Fig.~1) is due to the
destructive contribution of the VA interference contribution. Since
the VV+AA and VA contributions peak at different values of $T_\mu$
their having opposite signs leads to an oscillatory result.

\begin{figure}[ht]
\label{fig:ratio}
\includegraphics[scale=0.6, bb=90 150 469 750]{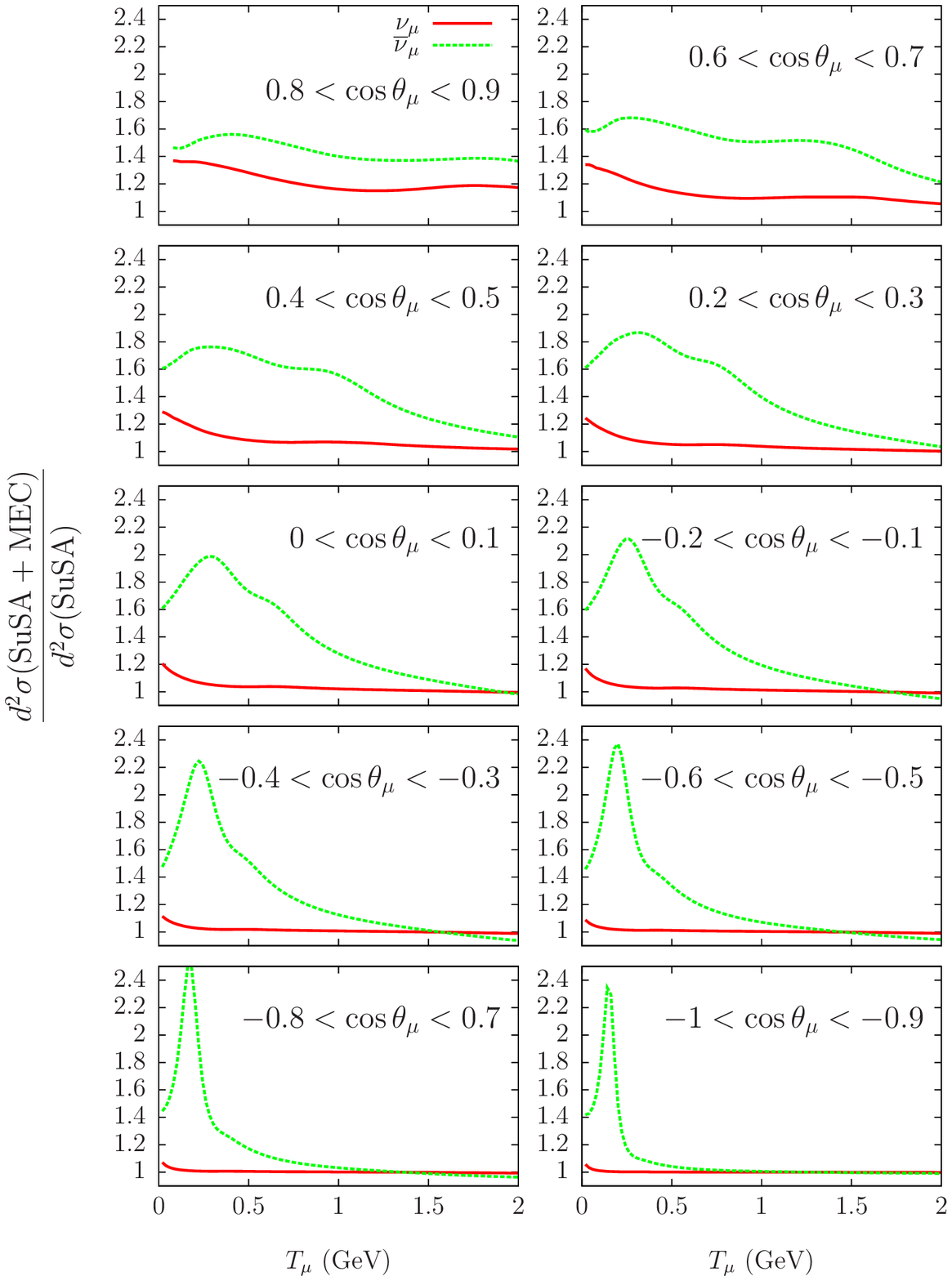}
\caption{ (Color online) Ratio of double-differential cross sections
  with and without MEC: $d^2\sigma(SuSA+MEC)/d^2\sigma(SuSA)$ for
  neutrino (lower curves, red) and antineutrino (upper curves, green)
  scattering versus muon kinetic energy $T_\mu$.}
\end{figure}

Note that in all of the figures discussed above we do not compare
with the most forward angles ($0.9<\cos\theta_\mu<1$), since for such
kinematics roughly 1/2 of the cross section arises from very low
excitation energies ($<$50 MeV), where the cross section is dominated
by collective excitations and any approach based on IA is bound to
fail.  This has been proven~\cite{Amaro:2010sd} for neutrinos and is
even more so for the MiniBooNE antineutrino case, where the
contribution of low excitation energies persists also at higher
scattering angles.  This is due to the fact that the mean energy of
the antineutrino flux at MiniBooNE is $\langle
E_{\overline\nu_\mu}\rangle\simeq$0.66 GeV, lower than the neutrino
one $\langle E_{\nu_\mu}\rangle\simeq 0.79$ GeV. Specifically, for the
lowest angular bin cutting out the contributions from the first 50 MeV
reduces the cross section by about 40\%, for the next bin ($0.8 < \cos
\theta_\mu < 0.9$) the reduction is about 18\%, and for larger angles
the reduction remains in the range 12--21\%, reaching a minimum for
$0.4 < \cos \theta_\mu < 0.5$, but rising again when going towards
180$^o$.

\begin{figure}[ht]
\label{fig:sigma}
\includegraphics[scale=0.75, bb=180 560 430 725]{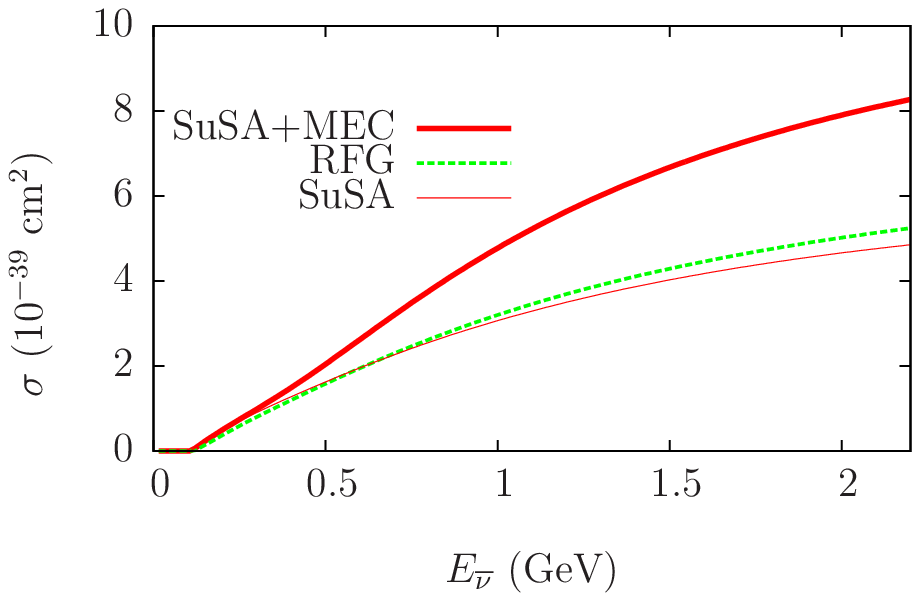}
\caption{ (Color online) Total cross section displayed versus 
antineutrino energy $E_{\overline\nu}$ for the pure 
SuSA model, the SuSA model plus MEC and the pure RFG model.}
\end{figure}
Finally, for completeness, in Fig.~4
we give the total cross section for antineutrino
scattering (see also \cite{Martini:2010ex}). 
Similar results for neutrinos using the present analysis were discussed in
\cite{Amaro:2011qb}.  While the SuSA and RFG QE results are rather
similar, it is obvious that the effect of MEC on the total cross
section is very large. For instance, at $E_{\overline\nu_\mu}= 1$ (2)
GeV this amounts to a 60 (70)\% increase. One warning should be
repeated here: what is displayed is $\sigma$ versus
$E_{\overline\nu_\mu}$, which can be done theoretically, whereas what
in experimental studies is called $E_{\overline\nu_\mu}^{QE}$ is 
something different.  
The latter is an effective energy obtained by assuming a model for where
the peak in the differential cross section occurs.

In summary, we have employed the SuperScaling Analysis together with MEC
in a study of antineutrino cross sections.
The effects of MEC in the double-differential cross sections are seen to be
very important, in fact, relative to the QE contribution, significantly
larger than for neutrinos.
Work is in progress to explore what happens when other models such as 
relativistic mean field theory are employed.

This work was partially supported by Spanish DGI and FEDER funds (FIS2008-01143, FIS2008-04189), by the Junta de Andalucia, by the Spanish Consolider-Ingenio 2000
programmed CPAN (CSD2007-00042), partly (TWD) by U.S. Department of Energy under cooperative agreement DE-FC02-94ER40818 and by the INFN-MICINN collaboration
agreement (ACI2009-1053).
%
%

\end{document}